# Air plasma key parameters for electromagnetic wave propagation at and out of thermal equilibrium: applications to electromagnetic compatibility

**P. Andre [1], G. Faure [1], A. Mahfouf [1], and S. Lalléchère [2]**

[1] LAEPT, EA 4646
Université Clermont Auvergne, Université Blaise Pascal, BP 10448, F-63000 Clermont-Ferrand, France
{pascal.andre}/{geraldine.faure}@univ-bpclermont.fr, mahfoufphysique@gmail.com

[2] Institut Pascal, UMR CNRS 6602
Université Clermont Auvergne, Université Blaise Pascal, BP 10448, F-63000 Clermont-Ferrand, France
sebastien.lallechere@univ-bpclermont.fr

***Abstract* —** This article addresses the importance of accurate characterization of plasma parameters for electromagnetic compatibility (EMC) purposes. Most of EMC issues involving plasma materials are obviously multi-physics problems (linking chemical, mechanical, thermal and electromagnetic wondering) with deep interactions. One of the main objectives of this paper is to establish the theoretical effect of thermal non-equilibrium on electromagnetic wave propagation. This will be characterized throughout plasma key parameters (including complex permittivity). Numerical simulations based upon Finite Integral Technique (FIT) will demonstrate the EMC interest of this methodology for shielding purposes and general air plasma.

## I. INTRODUCTION

Intentional or non-intentional plasma generations imply highly multi-physics studies involving chemistry, thermic, physics and of course electromagnetics to properly characterize electromagnetic (EM) fields. Previous studies [1-2] have demonstrated the benefit that could be taken from microwave breakdowns to improve shielding effectiveness (SE) of enclosures embedded with slots and equipment under test. Some current electromagnetic compatibility (EMC) issues require an accurate assessment of materials EM properties in various configurations: for instance damaging of aeronautical systems (wires, antennas) due to lightning, spacecraft re-entry (radio frequency, RF, plasma generation). Physical properties of such systems need some macroscopic data such as viscosity, electrical conductivity, internal energy. Thus, in several applications as discharges with liquid non-metallic electrodes, circuit breakers, arc tracking, RADAR applications, the electrons reach a temperature ($T_e$) higher than other chemical species ($T_h$). In this framework, the interaction between charged particles plays a key role. Computational electromagnetics requires efficient tools [3] in order to assess dielectric permittivity that depends greatly on the material and in the case of plasma on its thermodynamic state. Although thermal equilibrium (i.e. $\theta = T_e/T_h = 1$) is assumed in many test cases (decreasing computational needs), it is often badly used. Indeed, authors [4] have previously demonstrated that thermal non-equilibrium can play a major influence on argon plasma properties.

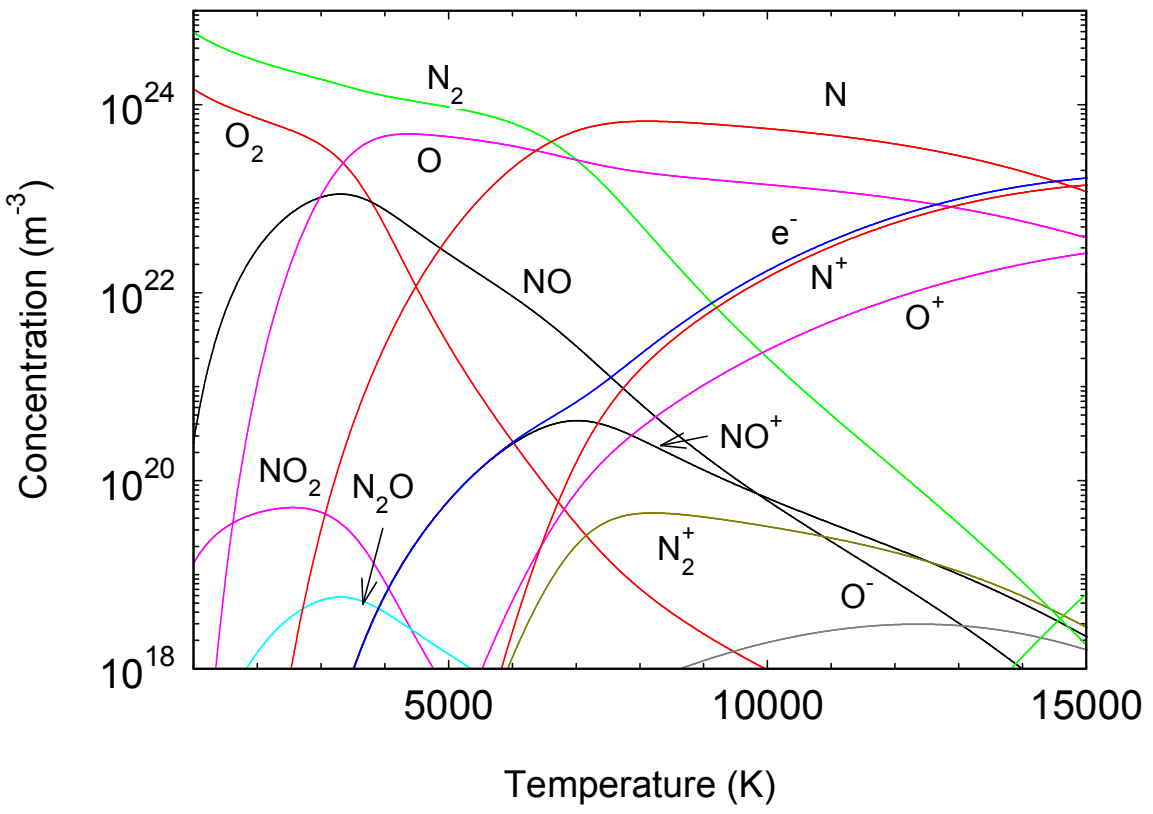

Fig. 1. Air composition at atmospheric pressure at thermal equilibrium.

Figures 1 and 2 clearly depict the high differences existing between concentrations of heavy species and electrons at and out of thermal equilibrium from Gibbs free energy minimisation [5] for air plasma at atmospheric pressure. This generalizes argon results with classical air composition (i.e. 80% $N_2$ and 20% $O_2$). It is to be noticed that ionisation appears at lower heavy species temperature when the thermal non equilibrium ratio $\theta$ increases for a given heavy species temperature $T_h$. Since EM properties highly depend on chemical concentrations, it involves potential huge effects on plasma behavioural modelling.

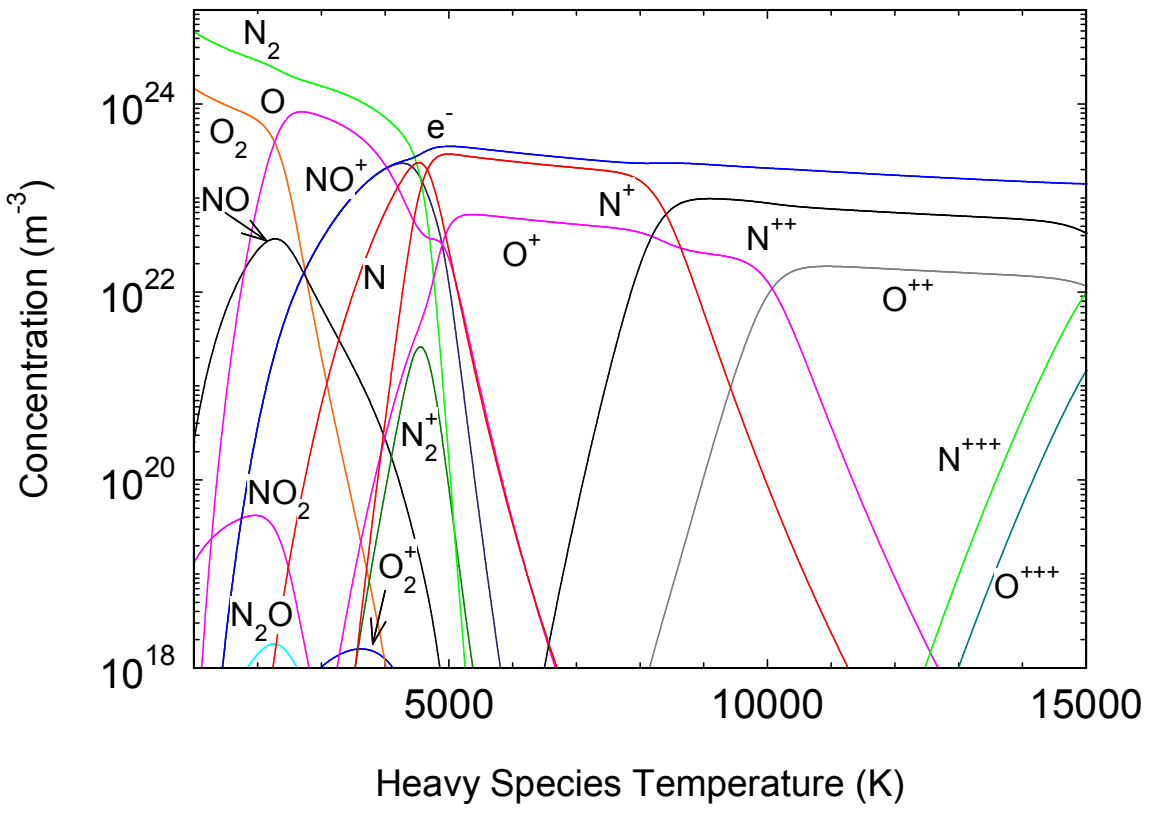

Fig. 2. Air composition at atmospheric pressure out of thermal equilibrium $\theta = T_e/T_h = 3$.

This article is organized as follows: section II briefly describes the theoretical methodology and plasma characteristic parameters are evaluated for several frequencies and air plasma in section; section III is dedicated to the electromagnetic impact of plasma modelling on field propagation, and an EMC illustrative example is proposed. The contribution ends with a conclusion and some prospects in section IV.

## II. PLASMA KEY PARAMETERS AT AND OUT OF THERMAL EQUILIBRIUM

### A. Theoretical model

By considering one electron inside a given electromagnetic environment (depicted by electric field $E$), one can obtain from the Newton's second law:

$$m_e \frac{dv}{dt} = -eE - k\, m_e v, \qquad (1)$$

where $-k\, m_e v$ is a restoring force, $v$ is the velocity of electron, and $m_e$ and $e$ are respectively mass and elementary charge of electron. Assuming the electric field as $\underline{E} = E_0 e^{i\omega t}$ and neglecting dipole creation inside plasma resolving (1), the celerity of electrons is obtained:

$$v = \frac{eE_0}{k+i\omega}\frac{1}{m_e} e^{i\omega t} \qquad (2)$$

Introducing the drift velocity one can obtain when $\omega = 0$ the parameter equal to the collision frequency $\upsilon_{e_p}$ of electrons with the other particles inside the plasma. So the real current density can be written as:

$$\vec{J_e} = -\frac{n_e e}{m_e}\left(\frac{-e\vec{E}_0}{\nu_{ep}+i\omega}\right) e^{i\omega t} \qquad (3)$$

Introducing effective current inside Ampere's law we obtain:

$$\vec{\nabla} \times \vec{H} = \varepsilon_0 \frac{\partial \vec{E}}{\partial t} + \vec{J_e} = \varepsilon \frac{\partial \vec{E}}{\partial t} \qquad (4)$$

Then the real permittivity is written as:

$$\varepsilon = \varepsilon_0 \left(1 - \frac{\omega_p^2}{\omega(\omega - i\,\upsilon_{ep})}\right). \qquad (5)$$

This permittivity is available for isotropic and non-magnetized plasma. We can feature key parameters: plasma pulsation $\omega_p = 2\pi f_p$ and collision frequency $\upsilon_{e_p}$ of electrons, and the thermal velocity of electrons. This last parameter will be given in the following, jointly with plasma characteristic parameters (i.e. $f_p$ and $\upsilon_{e_p}$) taking into account in an original way the physical properties of plasma material.

### B. Plasma characteristics and equivalent complex permittivity

Figure 3 shows frequency collisions of electrons and plasma frequency extracted from air composition and thermal assumptions (see Fig. 1-2). From equation (5), we can deduce that the formulation depends greatly on plasma state and wave frequency.

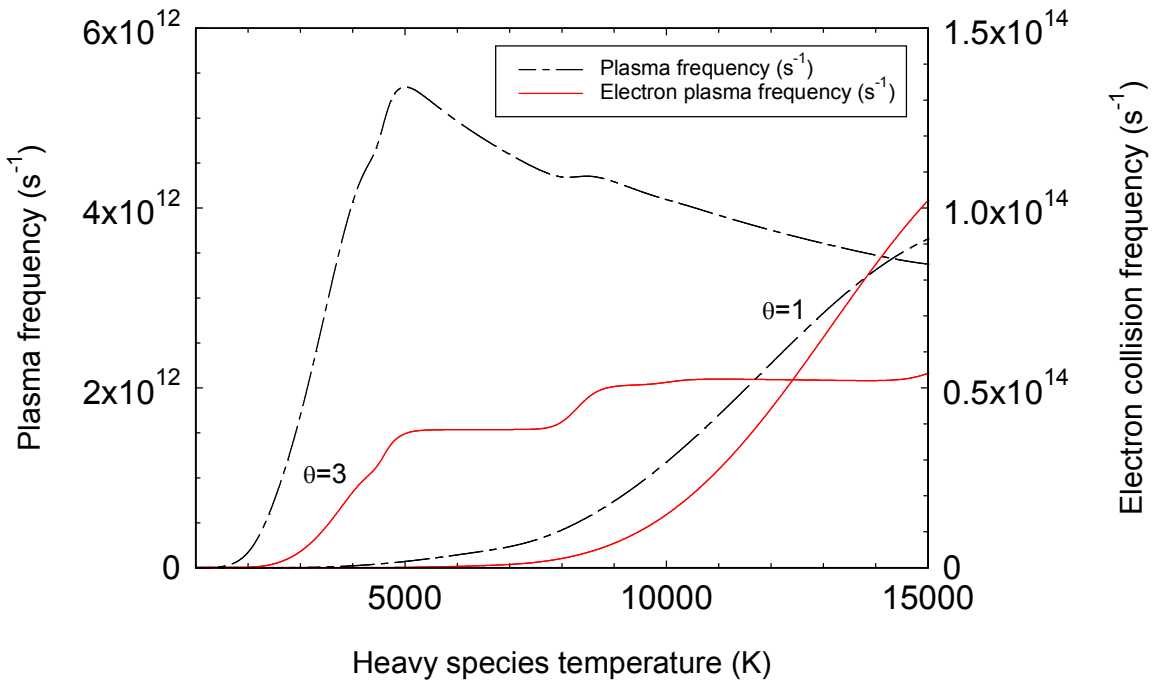

Fig. 3. Plasma frequency and electrons collisions frequency in air plasma at atmospheric pressure.

As a matter of fact, as we can see in Figure 3, the electron plasma frequency is restricted for weaker temperatures. As a first approximation, plasma can be considered as a dielectric material. In Figures 4-5, we have plotted the real and imaginary part of the relative permittivity for the air plasma at and out of thermal equilibrium. It is to be noted (data not shown here) that, for the lower temperature the real relative permittivity is close to 1 and the imaginary part is close to zero. Figures 4 and 5 show real components of dielectric constant are lower than unit ($T_h$ = 10,000 K).

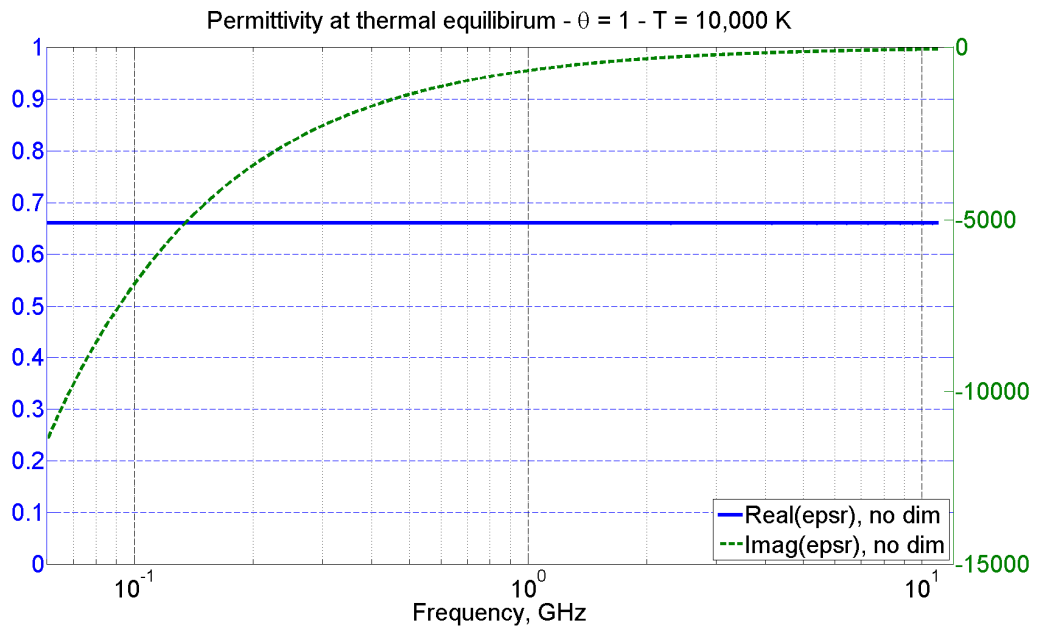

Fig. 4. Complex permittivity of the air plasma at thermal equilibrium ($\theta = 1$): real (blue) and imaginary (green) parts at $T_h$ = 10,000 K from 64 MHz to 10 GHz.

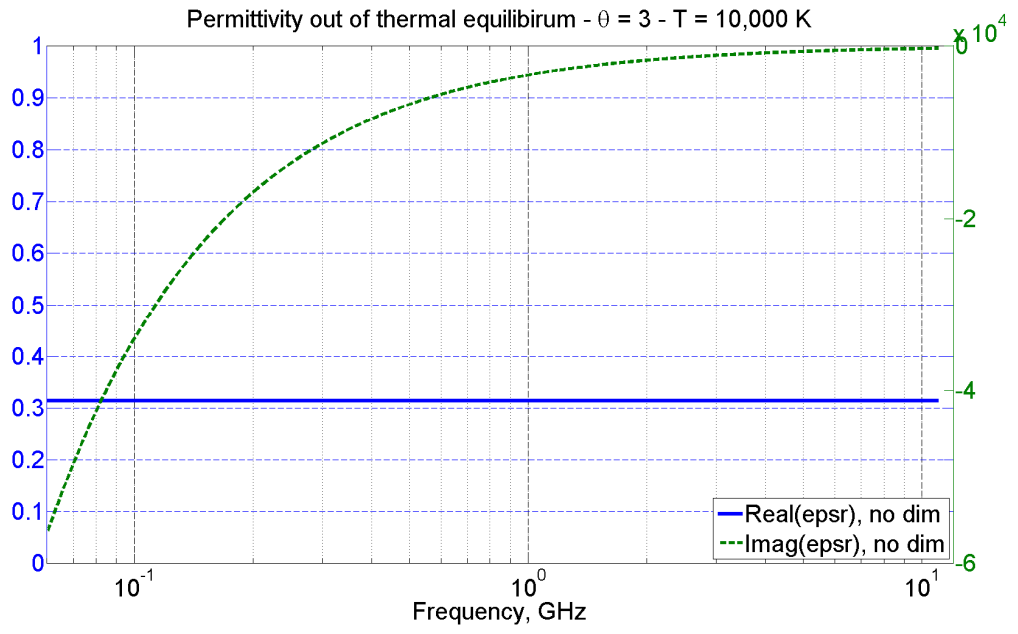

Fig. 5. Complex permittivity of the air plasma out of thermal equilibrium ($\theta = 3$): real (blue) and imaginary (green) parts at $T_h$ = 10,000 K from 64 MHz to 10 GHz.

## III. NUMERICAL RESULTS: PLASMA CHARACTERIZATION FOR SHIELDING EFFECTIVENESS (SE) APPLICATIONS

Some EM simulations were achieved using CST© MWS to assess EM field penetration inside air plasma at and without thermal equilibrium. Time solver and dispersive model based upon data from Figure 3 were used to quantify plasma shielding strength (i.e. E-field magnitude decreasing while penetrating plasma material) up to 5 GHz. For the sake of exhaustiveness, the next sections will detail numerical simulations.

### A. Test case #1: assessment of canonical shielding effectiveness (SE)

The first test case is inspired by works in [6]. Indeed, the aim of this paper was to characterize the influence of plasma physical parameters (e.g. concentrations and reactions between chemical species) on electromagnetic wave (EMW) propagation in to a slab. In this framework, the crucial part of the work relies on the characterization of material throughout models and plasma characteristics (i.e. plasma frequency $f_p$, and collision frequency $\upsilon_{e_p}$) as depicted in Figure 3. Those characteristics are highly dependent to the equilibrium (thermal and chemical composition) of heavy species and electrons.

The aims of this section are to demonstrate the difference that may be expected from taking into account (or not) potential thermal non-equilibrium jointly with the relevance of using computational electromagnetics tool (e.g. CST© with time domain solver). First of all, we put the focus on a canonical case, and in order to prepare numerical experiments in section III.2 (test case #2), we propose to model two kinds of plasmas (data given in Fig. 3) for $T_h$ = 10,000 K and $T_h$ = 15,000 K with CST© MWS at thermal and non-thermal equilibrium.

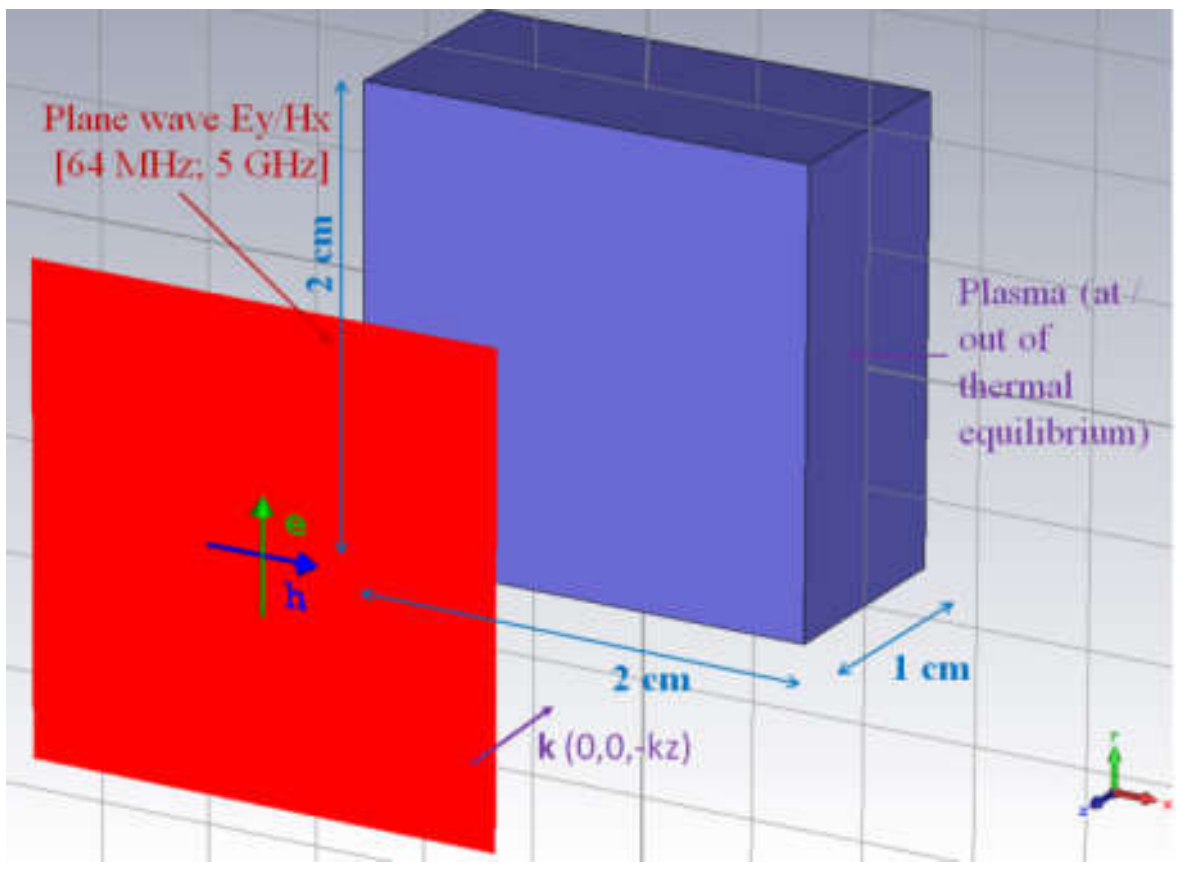

Fig. 6. Numerical setup for plane wave impinging on plasma slab (2 x 2 x 1cm$^3$, infinitely extended in x- and y- directions) using CST© MWS time solver and plasma material modelling (dispersive one).

Figure 6 depicts the numerical setup proposed for straightforward characterization of the EM attenuation of waves throughout plasma illuminated by a normal-incidence plane wave. The plasma slab is a 2x2x1 cm$^3$ volume (Fig. 6). The time simulation (CST© MWS, time solver) is maintained up to ensure at least that more than 40 dB of the maximum energy has vanished from the computational domain. The plasma dielectric dispersion relies on purely dispersive modelling according to data in Figures 4-5.

A huge number of potential EM applications of plasma layers exist in literature as expressed in the introduction. Canonical characterization of plasma attenuation at atmospheric pressure is carried out in [7] whereas spacecraft flight re-entry is studied in [6]. In each of the two previous cases, attenuation is defined in a different manner. We will next consider the shielding effectiveness (SE) of the plasma (Fig. 9) as follows:

$$SE = \frac{incident\ EMW}{transmitted\ EMW} \quad (6)$$

$$SE_{dB} = -20log\frac{E_{in}}{E_{out}} \quad (7)$$

where $E_{in}$ is the electric field located behind the infinite plasma slab (transmitted electromagnetic wave, EMW), and $E_{out}$ is the incident EMW.

In the following and based upon relations (6-7), the transmitted electric field ($E_{in}$) is computed from CST© time domain solver and dispersive medium given by original theoretical models from section II. The numerical results are compared to the analytical approach from [6] where the transmission coefficient $t$ ($t$=$E_{in}$/$E_{out}$ in relation (7)) is obtained as follows:

$$t = \frac{2\sqrt{\varepsilon_r}e^{ik_0d}}{2\sqrt{\varepsilon_r}cosh(ik_pd)+(\varepsilon_r+1)sinh(ik_pd)} \quad (8)$$

where $\varepsilon_r$ is the complex permittivity of plasma, $k_0$ is the wave number in bulk medium (air), $k_p$ is the wave number in plasma (here with different characteristics in terms of temperature, thermal equilibrium…), and $d$ is the width of considered plasma slab.

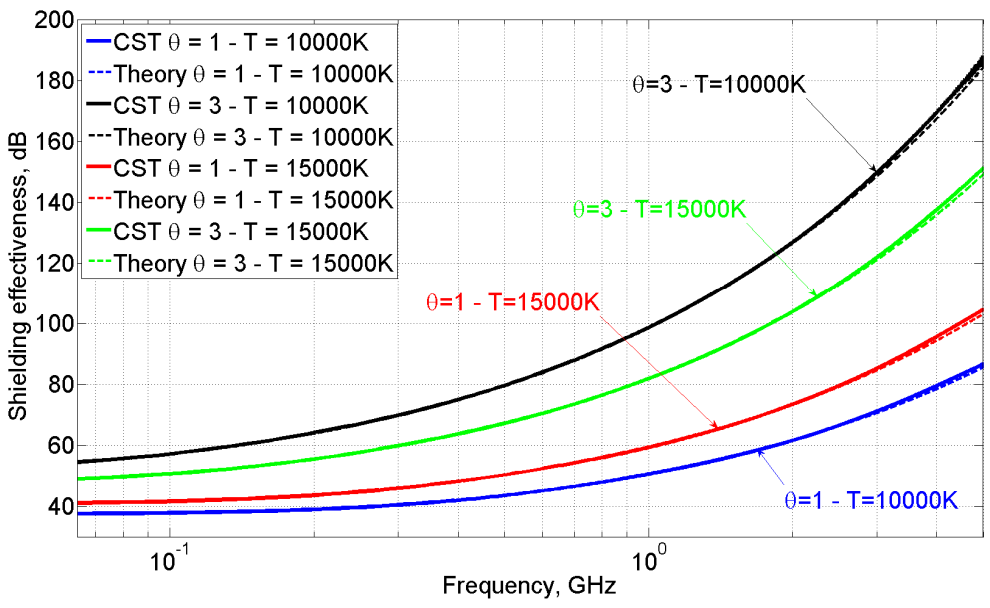

Fig. 7. Shielding effectiveness (attenuation in dB from 64 MHz to 5 GHz) at thermal equilibrium $\theta$=1 (blue, $T_h$ = 10,000 K; red, $T_h$ = 15,000 K) and out of thermal equilibrium $\theta$=3 (black, $T_h$ = 10,000 K; green $T_h$ = 15,000 K) relying on analytical formulation (dotted lines: reference [6]; dielectric permittivity from Fig. 4 and 5) and CST© (plain lines).

Figure 7 illustrates the impact of non-thermal equilibrium of air plasma at atmospheric pressure and at $T_h$ = 10,000 K / $T_h$ = 15,000 K on material SE in function of frequency; similarly to Figures 4-5, the dielectric properties of plasma are obtained at $T_h$ = 15,000 K (data not shown here) with original works based upon assessment of plasma characteristics from air plasma composition (Fig. 1-2). First, the results are in accordance with literature ones. Indeed, assuming similar characteristics of plasma (e.g. pressure, heavy species distribution, width of plasma slab), tens dB of attenuation are expected in [6-7]. Obviously, plasma composition and collision frequency (Fig. 1 and 2) play major roles, their increase leading to a proportional decrease of transmitted electric fields. The maximum gap existing between SE at and out of thermal equilibrium is higher for $T_h$ = 10,000 K than for $T_h$ = 15,000 K. Indeed, the gap is comprised between 3 dB and 45 dB considering heavy species temperature $T_h$ = 15,000 K, whereas SE is between 5 dB and 100 dB higher out of thermal equilibrium than at thermal equilibrium at temperature $T_h$ = 10,000 K. Finally, it is noted that the SE differences between $\theta$=3 and $\theta$=1 increases with frequency for each plasma temperature. Figure 7 validates use of fully dispersive plasma model given by the original theoretical model proposed in this work (section II). The next section will illustrate the importance of a careful definition of plasma characteristics (via complex permittivity and plasma characteristic frequencies) in EMC framework.

**B. Test case #2: cabinet shielding at and out of thermal equilibrium**

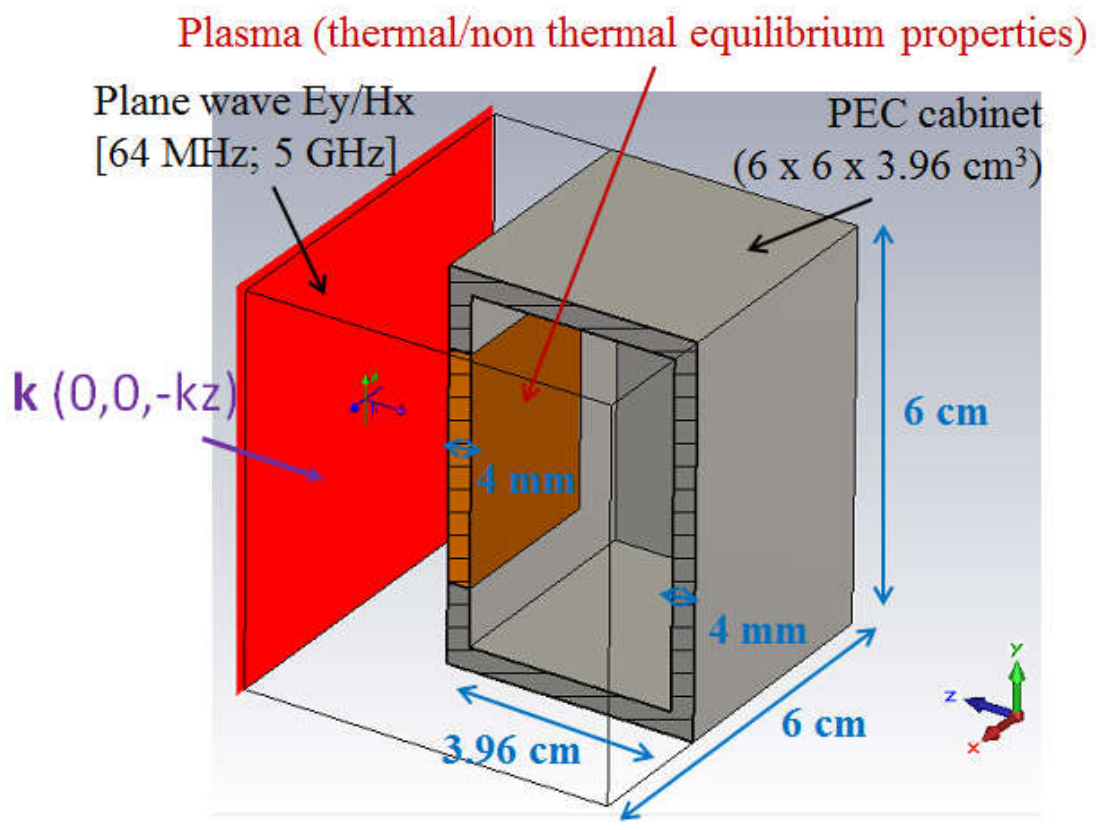

Fig. 8. Characterization of the influence of thermal ($\theta$=1) or non-thermal ($\theta$=3) equilibrium (numerical setup) on the SE of PEC cabinet (sectional view) subject to plane wave illumination.

Relying on previous results for canonical case (test case #1), we illustrate the influence of thermal or non-thermal equilibrium assumption throughout an EMC shielding example. The numerical configuration is illustrated in Figure 8: a perfectly conducting (PEC) enclosure (6 x 6 x 3.96 cm$^3$) is considered jointly with a square aperture (length=4 cm) and 4 mm-walls (and 4mm-width of plasma). Figure 8 shows the direction of the impinging plane wave (incident electric field $Ey = 260$ kV/m). Plasma characteristics are based upon results given in Figures 4-5 ($T_h = 10{,}000$ K).

Figure 9 represents the evolution of the SE of the cabinet in relation with frequency [0.064 MHz; 5 GHz]. The averaged gap existing between plasma at thermal (red) and non-thermal (green) equilibrium is between 9 dB and 45 dB. This perfectly justifies the EMC need for a particular care since plasma may reveal useful in that framework. It is to be noticed that, due to the proposed configuration (worst case regarding size of the aperture and plane wave source), the shielding effectiveness without plasma material quickly decrease within negative levels (i.e. field enhancement instead of shielding) from 2.8 GHz. Contrary to previous case, the plasma slab improves SE of the system (enclosure + plasma) up to 45 dB ($\theta$=1) and 85 dB ($\theta$=3). It should also be noticed that the system is subject to cavity resonances, decreasing SE for instance at $f$=4.859 GHz (resonance frequency in accordance with inner sizes of the enclosure). Finally, due to dispersive effect, plasma slab closes the cabinet and involve enhancement of 4.859 GHz-resonance frequency as depicted in Figure 9.

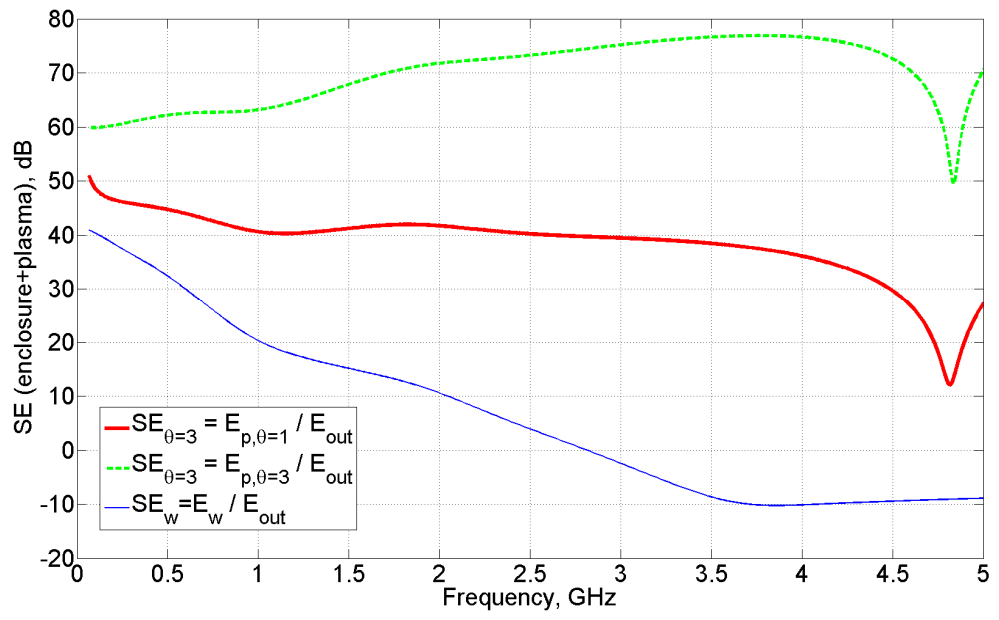

Fig. 9. SE of the cabinet (electric field measured at the center of the enclosure, Fig. 8) with plasma at (red) / out of (green) thermal equilibrium, and without plasma (blue, $E_w$) including only the enclosure. Results are given by normalizing data following relation (7) with $E_{out}$.

In order to illustrate the importance of taking into account inner thermal characteristics of plasma (differences between electrons and heavy species temperatures), it is proposed to focus on the influence of plasma by normalizing SE (Fig. 9) obtained in the two cases ($\theta$=1) and ($\theta$=3) by results computed without plasma material. Figure 10 shows the plasma attenuation (based upon electric field $Ey$-component computing) following the respective dB-differences $20log10(E_{\theta=1}/E_{out})$ and $20log10(E_{\theta=3}/E_{out})$ (see Fig. 9). As expected the cabinet is involved for a noticeable part in shielding characteristics. Figure 10 gives an overview of the dedicated effect of plasma material in the proposed EMC configuration (Fig. 8) by normalizing with test case involving only the enclosure. This lays emphasis on the importance of considering thermal equilibrium or not from theoretical model to EMC application since high gaps exist (from 10 dB to 40 dB) over the whole frequency bandwidth. As aforementioned in Figure 9 and due to the characteristics of starting resonance frequency (i.e. the presence of the air aperture, see Figure 8, involving both the resonance mode vanishing and a huge reflection of impinging plane wave), the shielding effectiveness is considerably spoiled in 'empty' case (without plasma, see blue line in Fig. 9). Contrary to previous case, plasma plays dual role since it affects the levels of fields penetrating in the cabinet but also closes it, enhancing first resonance mode influence around 4.859 GHz as illustrated in Figure 9.

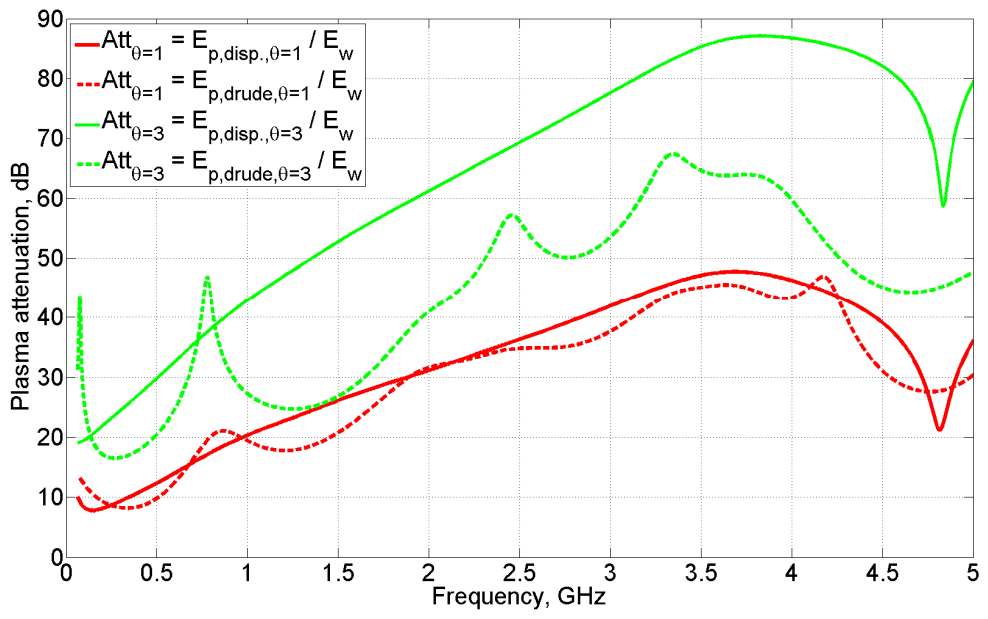

Fig. 10. Plasma attenuation including the effect of the enclosure (normalization with E-field given without plasma, $E_w$) at (red) and out of (green) thermal equilibrium; results are proposed including dispersive plasma medium (plain lines, see relation (5) and Fig. 4 and 5) and Drude's model (dotted lines) based upon plasma characteristics (Fig. 3).

Similarly to Figure 9, Figure 10 illustrates previous point and enriches the discussion by providing data obtained with time simulations taking into account the non-linear nature of plasma characteristics via Drude's modelling [3]. Indeed, in that case, the definition of plasma dielectric properties relies on intrinsic plasma characteristics (Fig. 3) jointly with a non-linear E-field threshold ("breakdown") modelling as explained in the following. By varying the plasma density within a given field level, plasma attenuation is decreased relatively to purely dispersive medium (see figures 7 and 9). It is to be noted in Figure 10 that weak gaps exist regarding thermal equilibrium ($\theta$=1, red curves); huger differences are obtained out of thermal equilibrium ($\theta$=3, green curves). The same trends between dispersive and Drude's material are observed in Figure 10 for $\theta$=1 (red) whereas up to 35 dB-gaps are computed for $\theta$=3 (green). These variations mostly depend on Drude's model including: plasma characteristics (e.g. plasma frequency $f_p$ and collision frequency $\upsilon_{e_p}$ from developments in section II, here see Fig. 3 and temperature $T_h$ = 10,000 K), and electric field breakdown level (here $E_{break}$ = 1 kV/m). As aforementioned in [8], plasma induced by microwave may efficiently offer EMC advantages by providing interesting EM shielding properties. In this case, we demonstrate the capability of time domain simulations (including Drude's modelling and breakdown level) to enrich purely dispersive approach. This also lays emphasis on the huge importance of properly defining plasma characteristics (plasma frequencies and/or complex permittivity) in EMC context, especially when thermal equilibrium assumption is not satisfied. It should be noticed that, for air plasma at atmospheric pressure, purely dispersive plasma modelling is sufficient to accurately assess the EM shielding properties of the material. In this framework, plasma characterization may be useful to improve the assessment of EMC shielding.

## IV. CONCLUSION AND PROSPECTS

This contribution aims at demonstrating the importance of modelling plasma behaviour in EMC framework. In this context, a particular care needs to be taken in order to properly define the impact of the physical conditions assumed for the definition of plasma. Of course, it is well-known that the composition, temperature, pressure of the material (plasma) is of great importance. The thermal equilibrium respectively between the temperatures of electrons and heavy species plays also a key role as illustrated by the different characteristics of plasma (i.e. plasma and collision frequencies) given at and out of thermal equilibrium. Obviously, this involves major changes regarding the dielectric properties of the material (complex permittivity; non-thermal equilibrium may lead to increase dielectric losses up to a scaling factor of 6 in comparison with thermal equilibrium assumption).

In this paper, we have shown how this original plasma modelling enriches the understanding of shielding properties in EMC context. Thus, we have observed comparable levels of electromagnetic attenuation than results found in literature at thermal equilibrium. On the contrary, non-thermal equilibrium may involve noticeable differences (here 40 dB at maximum). Using "Full-Wave" simulation tool such as CST© jointly with the proposed theoretical plasma models enrich the physical understanding of wave propagation in complex media. Moreover, the assessment of EMC criteria (e.g. shielding effectiveness) is improved.

Further works are nowadays under consideration to enhance this study. Parametric and multi-physics works based upon these models may be useful for EMC applications and/or various electromagnetic issues (e.g. material characterization, plasma, lightning, transport, space re-entry, and communications). It should also be useful to assess the effect of non-linear field behaviour due to plasma inclusion. Based upon proposed work, it should be noticed that plasma material may be modelled throughout use of proposed plasma frequency and collision frequency (parallel to complex permittivity). This may lead to enrich time domain model and illustrates threshold effects in EMC context (involving shielding or field enhancement) and offers an extension to multi-physics issues (e.g. electromagnetic and thermal ones).